# Building Reliable Budget-Based Binary-State Networks

Wei-Chang Yeh
Integration and Collaboration Laboratory
Department of Industrial Engineering and Engineering Management
National Tsing Hua University
yeh@ieee.org

*Abstract* — Everyday life is driven by various network, such as supply chains for distributing raw materials, semi-finished product goods, and final products; Internet of Things (IoT) for connecting and exchanging data; utility networks for transmitting fuel, power, water, electricity, and 4G/5G; and social networks for sharing information and connections. The binary-state network is a basic network, where the state of each component is either success or failure, i.e., the binary-state. Network reliability plays an important role in evaluating the performance of network planning, design, and management. Because more networks are being set up in the real world currently, there is a need for their reliability. It is necessary to build a reliable network within a limited budget. However, existing studies are focused on the budget limit for each minimal path (MP) in networks without considering the total budget of the entire network. We propose a novel concept to consider how to build a more reliable binary-state network under the budget limit. In addition, we propose an algorithm based on the binary-addition-tree algorithm (BAT) and stepwise vectors to solve the problem efficiently.

## 1. INTRODUCTION

Networks have several practical characteristics, including universality, simplicity, and versatility. Hence, more and more real-world systems are being constructed, planned, designed, executed, managed, and controlled using networks for transmitting, transferring, and/or transporting petrol [1], gas [2], water [3], power [4], products [5], signals [6], data [7], vehicles [8], multimedia [9], social relationships [10], etc.

To get a better and more convenient understanding of a network, we need to consistently manage the network and evaluate its performance [11, 12, 13]. Network reliability is the probability that the current network functions, performs, or executes tasks successfully. Hence, network reliability has been a common indicator for evaluating the functions, states, and performances of various networks for many decades [11, 12, 13].

A binary-state network has a simple structure such that all its components are in the binary-state: working or failed [14]. All types of networks are constructed based on a binary-state network [15, 16]. Multistate flow networks (MFNs) allow each component to have multiple states and follow the flow conservation law [15, 16, 17, 18]. Based on the MFN, multistate information networks (MINs) further discard the flow conservation law. Multi-commodity networks, including multi-commodity MFNs (MMFNs) [19] and multi-commodity MINs (MMINs) [20] are generalized from MFN and MIN to accept different types of flows, called commodities, in the network to meet real-life applications for some current networks, such as cloud computing, multimedia, and social networks. Multidistribution networks further extend multicommodity networks to allow different commodities in each component to have different state distributions [21].

Regardless of whether the component in the network is binary-state or multistate, single or multi-commodity, and satisfies or dissatisfies the flow conservation law, with one or many state distributions, it is NP-Hard and # P-hard to calculate the network reliability [12, 13]. Different approximate methods, for example, simulations [22], bounds [23], and AI [14], have been proposed to overcome the NP-Hard obstacles to evaluating the approximated performances of practical larger networks.

However, owing to the advancement of some novel concepts [24], the calculation of exact reliability is more efficient than ever. For example, for the redundancy allocation problem, the exact-reliability algorithm based on binary-addition-tree (BAT) proposed in [24] outperforms the best AI based algorithm on simplified swarm optimization (SSO) [25], which needs to be implemented more than 30 times to obtain a robust result. Moreover, the parallel network reliability computing using multithread CPU and GPU increases the efficiency of some exact-reliability algorithms [26].

A binary-state network is the root of numerous networks. The improvement in calculating binary-state network reliability thus also enhances the efficiency of calculating the reliability of all types of networks. Thus, this study focuses on improving the calculation of the exact reliability of binary-state networks.

Moreover, various limitations and new constraints occur in real-life networks, such as cost [15, 17, 28], weight [24, 25], $k$-out-of-$n$ [27], transmission speed [16], volume [24, 25], memory capacity, signal quality [6], and runtime [16]. Therefore, more attention has shifted to conditional network reliability problems to address real limitations, such that conditional network reliability problems are more important than these problems without accounting for constraints.

All conditional binary-state network reliability algorithms, including the MP-based budget limit problems [15, 17, 28, 29, 30, 31], are either based on minimal paths (MPs) [15, 17, 28, 29, 30] or minimal cuts (MCs) [31]. Both MPs and MCs are arc subsets. Each arc in MPs or MCs is not redundant and none of them can be removed from MPs or MCs. However, existing studies on MP-based budget limit problems are focused on each minimal path (MP) and lack consideration of the total budget of the entire network [15, 17, 28, 29, 30]. A novel budget limit problem is proposed to build a more reliable binary-state network under the budget limit for all networks. Note that finding all MPs or MCs is NP-Hard and #P-hard [12, 13].

Another purpose of this study is to propose a novel exact-solution algorithm for the novel budget-limited binary-state network reliability problem to increase network reliability under budget limitations. The proposed conditional binary-state network reliability problem is completely different from the existing one, which only considers the budget of each identified MP. Considering the budget of the entire network rather than the budget of each MP, the proposed problem is more applicable and reasonable.

The remainder of this paper is organized as follows. All acronyms, notations, nomenclatures, and assumptions for the proposed novel problem and novel algorithm are defined in Section 2. A short review of the current MP-based algorithms, BAT, and path-based layer-search algorithm (PLSA) is presented in Section 3. Section 4 introduces the proposed budget-limit network reliability problem and the MP-based budget limit problem, together with a discussion of their differences. The details of the major innovations in the proposed algorithm are presented in Section 5. The pseudocode, time complexity, and demonstration of the proposed algorithm are presented in Section 6. Section 7 concludes the study with remarks and discusses possible future work.

## 2. ACRONYMS, NOTATIONS, NOMENCLATURE, AND ASSUMPTIONS

All necessary acronyms, notations, nomenclature, and assumptions are presented here.

### 2.1 Acronyms

- MC: Minimal cut
- MP: Minimal path
- BAT: Binary-addition tree algorithm [34]
- PLSA: Path-based layered-search algorithm
- IET  Inclusion-exclusion technology
- SDP: Sum-of-disjoint product

### 2.2 Notations

- $|\bullet|$: Number of elements in set $\bullet$
- $n$: Number of nodes
- $m$: Number of arcs
- $V$: Set of nodes $V = \{1, 2, \ldots, n\}$, for example, $V = \{1, 2, 3, 4\}$ in Figure 1.

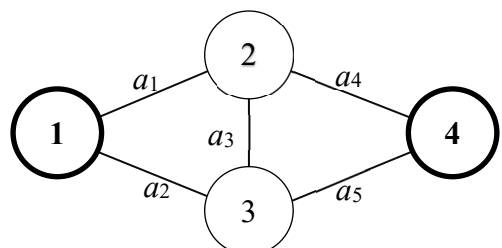

**Figure 1.** Example network.

$E$: Set of arcs $E = \{a_1, a_2, \ldots, a_m\}$, e.g., $E = \{a_1, a_2, \ldots, a_5\}$ in Figure 1.

$a_i$: $i^{th}$ arc in $E$

$a_{[k]}$: $[k]^{th}$ arc in $E$ such that $a_{[k]} \leq a_{[k+1]}$ for $k = 1, 2, \ldots, (m-1)$

$e_{u,v}$: directed arc from nodes $u$ to $v$ in $E$

$G(V, E)$: Graph with $V$, $E$, source node 1, and sink node $n$; e.g., Figure 1 is a graph with source node 1, and sink node 4.

$C^*$: Budget limit

$C(a)$: Cost to have $a$.

$\Pr(a)$: Probability that $a$ is still working.

$\mathbf{D_b}$: $\mathbf{D_b} = \{(\Pr(a), C(a)) \mid \text{for all } a \in E\}$.

$G(V, E, \mathbf{D_b})$: Network with $G(V, E)$ and $\mathbf{D_b}$.

$x_i$: State of $a_i$ such that $x_i = 0$ and 1 if $a_i$ is failed and working, respectively, for $i = 1, 2, \ldots, m$.

$X$: $m$-tuple binary-state vector $X = (x_1, x_2, \ldots, x_m)$.

$X(a_i)$: $X(a_i) = x_i$ for $i = 1, 2, \ldots, m$.

$G(X)$: $G(X) = G(V, X) \subseteq G(V, E)$ related to $X$ such that $a_i \in G(X)$ and $a_i \notin G(X)$ if $x_i = 1$ and 0, respectively. For example, $G(X)$ is shown in Figure 2, where $X = (1, 1, 0, 1, 1)$, $G(V, E)$ is provided in Figure 1, and arcs $a_3$ shown in dashed line is failed.

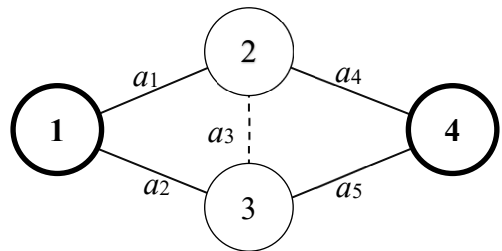

**Figure 2.** $G(X)$ in Figure 1, where $X = (1, 1, 0, 1, 1)$.

$C(X)$: $C(X) = x_1 C(a_1) + x_2 C(a_2) + \ldots + x_m C(a_m)$ is the cost to build $G(X)$.

$S$: stepwise vector $S = (s_1, s_2, \ldots, s_\mu)$.

$B(S)$:   $m$-tuple binary-state vector corresponding to the stepwise vector.

$\Pr(X)$:   $\Pr(X) = \prod_{X(a)=1} \Pr(a) \times \prod_{X(a)=0} (1 - \Pr(a))$  for all $a \in E$.

$R(X)$:   Reliability of $G(X)$.

$\Pr(p)$:   $\Pr(p) = \prod_{a \in p} \Pr(a)$  for all MP $p$.

$\pi$:   Number of MPs

$\pi^*$:   Number of feasible MPs

$L_i$:   Layer $L_i = \{v \mid e_{u,v} \in E,$ for all $u \in L_{(i-1)}$ and $v \notin (L_1 \cup L_2 \cup \ldots \cup L_{(i-1)})\}$.

$m_{\max}$:   Maximum number of working arcs in each feasible $X$

$m_{\min}$:   Minimum number of working arcs in each feasible $X$

$B_d^=$:   $B_d^= = \{X = (x_1, x_2, \ldots, x_m) \mid$ for all vectors $X$ obtained from BAT with $\sum_{k=1}^{m} C(a_k) = d$ and $X$ is not less than $X\hat{\ }$ for all feasible $X\hat{\ } \in B_{d^*}^=$ and $d < d^* \leq m_{\max}\}$

$X_d^*$:   Best feasible vector among all feasible vectors in $B_d^=$

$X_d^\#$:   Best feasible vector among all vectors in $B_d^=$

## 2.3 Nomenclature

Reliability:   The probability that there is at least one path from node 1 to node $n$.

Redundant arc:   Arc $a$ is redundant if its removal has no effect on the functions of the original arc subset, such as paths, cuts, and the connected graphs. For example, $a_3$ and $a_5$ are redundant in path $A = \{a_2, a_3, a_5\}$ and cut $B = \{a_1, a_2, a_5\}$, respectively, as illustrated in Figure 1.

MC:   A special arc subset without redundant arcs, such that its removal disconnects nodes 1 and $n$. For example, $\{a_1, a_3, a_5\}$ is an MC, as illustrated in Figure 1.

MP:   A simple path such that nodes 1 and $n$ are connected. For example, $\{a_1, a_3, a_5\}$ is an MP, as illustrated in Figure 1.

Connected graph:   $G(X)$ is connected if each node and arc in $G(X)$ can be connected via at least one directed path from node 1 to $n$.

Connected vector:  Vector $X$ is connected if $G(X)$ is connected.

Feasible (connected) Vector:   Vector $X$ is feasible if $X$ is connected and $C(X) \leq C^*$.

### 2.4 Assumptions

1. $G(V, E)$ is connected without loops or parallel arcs.
2. Each arc state is binary with statistically independent probability.
3. Each node is perfectly reliable.

## 3. OVERVIEW OF MP, BAT, and PLSA

The background for the proposed problem and algorithm, including the MP, BAT, and PLSA, is introduced here.

### 3.1 MP and MP-Based Algorithms

An MP is a simple path, and none of its arcs is redundant, i.e., the removal of any arc causes the remaining MP to not be a path. For example, there are only four MPs, $p_1 = \{a_1, a_4\}$, $p_2 = \{a_1, a_3, a_5\}$, $p_3 = \{a_2, a_5\}$, and $p_4 = \{a_2, a_3, a_4\}$, as shown in Figure 1. Both arc subsets $\{a_1, a_3\}$ and $\{a_1, a_3, a_4, a_5\}$ are not MPs, because $\{a_1, a_3\}$ is not a path from nodes 1 to 4; $\{a_1, a_3, a_4, a_5\}$ is a path connecting nodes 1 and 4, but $a_4$ is redundant and can be discarded from $\{a_1, a_3, a_4, a_5\}$ without disconnecting nodes 1 and 4.

After having all MPs, an extra algorithm, for example, the inclusion-exclusion algorithm [32] or the sum-of-disjoint product algorithm [32], is needed to calculate the final reliability in terms of MPs. Hence, MP algorithms that find all MPs are indirect algorithms. However, MPs are still very important for calculating the exact reliability of conditional binary-state network reliability [15, 17, 28, 29, 30]. Therefore, fruitful MP algorithms have been proposed based on implicit enumeration algorithms and can be categorized as universal generating function methodology [34], breadth-search-first [34], depth-search-first [35], and heuristic algorithm [36].

## 3.2 BAT

Yeh first proposed BAT based on the following two rules to efficiently produce all possible binary-state vectors [37]:

**Rule 1.** Find the first zero coordinate, say, $x_i$, and let $x_j = 0$ for $j < i$.

**Rule 2.** If $i$ is the last coordinate, halt, and all vectors are found.

With the above two simple rules, the pseudocode of BAT is described as follows [37, 38].

**Algorithm 1: BAT** [37, 38]

**Input:** The value of the $k$ in finding all $k$-tuple binary-state vectors.

**Output:** All $k$-tuple binary-state vectors.

**STEP 0.** Let $X = (x_1, x_2, \ldots, x_k) = \mathbf{0}$ and $i = 1$.

**STEP 1.** If $x_i = 0$, let $x_i = 1$, $i = 1$, and go to STEP 1.

**STEP 2.** Halt, if $i$ is the last coordinate, i.e., $i = k$.

**STEP 3.** Let $x_i = 0$, $i = i + 1$, and go to STEP 1.

From the above, BAT has four steps and one $k$-tuple binary-state vector $X$, which is repeatedly updated. Hence, BAT is simple to code, efficient in operation, economical in computer memory, and flexible to make-to-fit. Thus, numerous different applications have been applied to BAT in calculating network reliability problems [22, 37, 38], resilience appraisal [39], wildfire spread probability [40], social networks [10], and computer virus propagation probability [41].

For example, all 5-tuple binary-state vectors $X$ obtained from Figure 1 using the above 4-step BAT pseudocode are shown in Table 1, where the subscription $i$ of $X$ is unnecessary in the BAT, and readers just need to recognize which iteration is necessary to obtain the corresponding $X$ easily. The last column indicates whether $X$ is a real MP based on the degree of each arc in $X$.

**Table 1.** All obtained 5-tuple vectors and MPs using the BAT.

| $i$ | $X_i$ | MP? | $i$ | $X_i$ | MP? |
|---|---|---|---|---|---|
| 1 | (0, 0, 0, 0, 0) | | 17 | (0, 0, 0, 0, 1) | |
| 2 | (1, 0, 0, 0, 0) | | 18 | (1, 0, 0, 0, 1) | |
| 3 | (0, 1, 0, 0, 0) | | 19 | (0, 1, 0, 0, 1) | Y |
| 4 | (1, 1, 0, 0, 0) | | 20 | (1, 1, 0, 0, 1) | |
| 5 | (0, 0, 1, 0, 0) | | 21 | (0, 0, 1, 0, 1) | |
| 6 | (1, 0, 1, 0, 0) | | 22 | (1, 0, 1, 0, 1) | Y |
| 7 | (0, 1, 1, 0, 0) | | 23 | (0, 1, 1, 0, 1) | |
| 8 | (1, 1, 1, 0, 0) | | 24 | (1, 1, 1, 0, 1) | |
| 9 | (0, 0, 0, 1, 0) | | 25 | (0, 0, 0, 1, 1) | |
| 10 | (1, 0, 0, 1, 0) | Y | 26 | (1, 0, 0, 1, 1) | |
| 11 | (0, 1, 0, 1, 0) | | 27 | (0, 1, 0, 1, 1) | |
| 12 | (1, 1, 0, 1, 0) | | 28 | (1, 1, 0, 1, 1) | |
| 13 | (0, 0, 1, 1, 0) | | 29 | (0, 0, 1, 1, 1) | |
| 14 | (1, 0, 1, 1, 0) | | 30 | (1, 0, 1, 1, 1) | |
| 15 | (0, 1, 1, 1, 0) | Y | 31 | (0, 1, 1, 1, 1) | |
| 16 | (1, 1, 1, 1, 0) | | 32 | (1, 1, 1, 1, 1) | |

## 3.3 Layers and PLSA

Network reliability is the probability that nodes 1 and $n$ are connected successfully under redefined conditions, such as the budget limit here. Hence, verifying whether nodes 1 and $n$ are connected is important in network reliability problems.

The PLSA modified from the layer-search algorithm [42], is very efficient for checking the connectivity of nodes 1 and $n$ in BATs. Hence, PLSA was also adapted to verify the connections between nodes 1 and $n$ in the proposed algorithm.

PLSA builds disjoint node subsets $L_1$, $L_2$, …, $L_l$ consecutively until there is no other layer existing or node $n$ is already included in the last found layer, where $L_i = \{v \mid e_{u,v} \in E,$ for all $u \in L_{(i-1)}$ and $v \notin (L_1 \cup L_2 \cup … \cup L_{(i-1)})\}$ is called the layer in [42]. If $n \in L_l$, nodes 1 and $n$ are connected; otherwise, no connection exists between nodes 1 and $n$. The pseudocode of the PLSA is as follows.

**Algorithm 2: PLSA** [42]

**Input:**  $G(V, X)$.

**Output:**  Whether $X$ is connected, i.e., there is a directed path from nodes 1 to $n$ in $G(X)$.

**STEP 0.**  Let $i = 2$ and $L = L_1 = \{1\}$.

**STEP 1.**  Let $L_i = \{v \notin L \mid$ for all $e_{u,v} \in E$ and $u \in L_{i-1}\}$.

**STEP 2.** If $n \in L_i$, stop and $X$ is connected.

**STEP 3.** If $L_i = \emptyset$, stop and $X$ is connected.

**STEP 4.** Let $L = L \cup L_i$, $i = i + 1$, and return to STEP 1.

From the definition and the above pseudocode, $(L_1 \cup L_2 \cup \ldots \cup L_l) \subseteq V$ and $(L_i \cup L_j) = \emptyset$ for all $i \neq j$. Hence, PLSA is straightforward and efficient with a time complexity of only $O(n)$. PLSA was implemented in the proposed algorithm to verify the connectivity of each found $X$.

Consider the graph in Figure 2 as an example, where $X = (1, 1, 0, 1, 1)$. From the pseudocode of PLSA, these iterations verify the connectivity of $G(X)$ as listed in Table 2.

**Table 2.** Process from the proposed PLSA for Figure 2.

| $i$ | $L_i$ | $L$ | Remark |
| --- | --- | --- | --- |
| 1 | {1} | {1} | Node 1 is the source node |
| 2 | {2, 3} | {1, 2, 3} | $e_{1,2}, e_{1,3} \in E$ |
| 3 | {4} | | Node 4 is the sink node |

## 4. BUDGET-LIMITED NETWORK RELIABILITY PROBLEM

The proposed budget-based network reliability problem is discussed in this section, along with a comparison of the major differences between the proposed budget-based network reliability problem and the original MP-based budget-limited network reliability problem.

### 4.1 MP-Based Problem

The traditional MP-based budget-limited network reliability problem requires finding all MPs: $p_1, p_2, \ldots, p_\pi$, such that

$$C(p_i) = \sum_{\forall a \in p_i} C(a) \leq C_{ub}. \tag{1}$$

Note that the above equation only considers the cost of each MP rather than the total cost, which may exceed the budget limit.

After filtering out and discarding the MPs with cost over $C_{ub}$, the remaining feasible MPs, such as $p_1, p_2, \ldots, p_{\pi^*}$, are used to calculate the MP-based budget-limit binary-state network reliability in terms of the inclusion-exclusion technology [32] or the sum-of-disjoint product algorithms [33] based on the following two equations, respectively:

$$\Pr(\bigcup_{i=1}^{\pi^*} p_i) = \Pr(\{X \mid p_1 \leq X\}) + \Pr(\{X \mid p_2 \leq X \text{ and } X < p\}) + \ldots + \Pr(\{X \mid p_{\pi^*} \leq X, X < p_1, \ldots, X < p_{\pi^*-1}\}). \quad (2)$$

where $p_k$ is written in the vector form $(p_{k,1}, p_{k,2}, \ldots, p_{k,m})$ such that $p_{k,i} = \begin{cases} 1 & \text{if } a_i \in p_k \\ 0 & \text{otherwise} \end{cases}$ in Eq. (2).

For example, there are four MPs in Figure 1: $p_1 = \{a_1, a_4\}$, $p_2 = \{a_1, a_3, a_5\}$, $p_3 = \{a_2, a_5\}$, and $p_4 = \{a_2, a_3, a_4\}$. Let the budget limit be equal to 14. The cost and reliability of each arc are listed in Table 3.

Table 3. Cost and reliability of each arc in Figure 1.

| $i$ | 1 | 2 | 3 | 4 | 5 |
|---|---|---|---|---|---|
| $C(a_i)$ | 2 | 4 | 6 | 7 | 10 |
| $\Pr(a_i)$ | 0.95 | 0.90 | 0.85 | 0.80 | 0.75 |

From Table 3, Eq. (2), and

$$\Pr(p_i) = \sum_{\forall a \in p_i} \Pr(a), \quad (3)$$

we have the cost and reliability of each MP listed in Table 4.

Table 4. Cost and reliability of each MP in Figure 1.

| $i$ | 1 | 2 | 3 | 4 |
|---|---|---|---|---|
| $p_i$ | $\{a_1, a_4\}$ | $\{a_1, a_3, a_5\}$ | $\{a_2, a_5\}$ | $\{a_2, a_3, a_4\}$ |
| $C(p_i)$ | 9 | 18(>14) | 14 | 17(>14) |
| $\Pr(p_i)$ | 0.76 | 0.605625 | 0.675 | 0.612 |

The costs of both $p_2$ and $p_4$ are over the budget limit 14, $14 < C(p_2) = 18$, and $14 < C(p_4) = 17$, and must be discarded. The remaining MPs, $P_1$ and $P_3$, are used to calculate the conditional binary-state reliability from the inclusion-exclusion method as follows: if $\Pr(a_1) = 0.95$, then $\Pr(a_2) = 0.90$, $\Pr(a_3) = 0.85$, $\Pr(a_4) = 0.80$, and $\Pr(a_5) = 0.75$.

$$R(\{p_1 \cup p_3\}) = \Pr(p_1) + \Pr(p_3) - \Pr(p_1 \cap p_3)$$

$$= \Pr(\{a_1, a_4\}) + \Pr(\{a_2, a_5\}) - \Pr(\{a_1, a_2, a_4, a_5\})$$

$$= 0.95 \times 0.80 + 0.90 \times 0.75 - 0.95 \times 0.80 \times 0.90 \times 0.75$$

$$= 0.922. \tag{4}$$

As shown in Figure 2, four arcs, $a_1$, $a_2$, $a_4$, $a_5$, were used to construct the final network and calculate reliability. The total cost of having these four arcs is $C(a_1) + C(a_2) + C(a_4) + C(a_5) = 23$, which is much larger than the budget limit, i.e., it is not reasonable to consider the cost of each MP only. Hence, there is a need to fix the above problem to apply the budget-limited network reliability problem in real-life conditions.

### 4.2 Proposed Problem

The proposed budget-based binary-state network reliability problem is discussed as follows: in the proposed problem, the budget is for the entire network, rather than for each MP. Hence, the proposed problem must find the best subnetwork with maximal reliability that satisfies the budget limit. Let $X_i = (x_{i,1}, x_{i,2}, \ldots, x_{i,\pi_i})$ be an arc vector such that

$$x_{i,j} = \begin{cases} 0 & \text{if } a_i \notin G(X_i) \\ 1 & \text{otherwise} \end{cases}. \tag{5}$$

The cost in building the related subnetwork $G(X_i)$ is defined below:

$$C(X_i) = \sum_{\forall X_i(a)=1} C(a) \leq C^*. \tag{6}$$

Consider the example discussed in Section 3.2 but with the new budget limit of 26. The corresponding 5-tuple vector and its costs are listed in the 2nd and 3rd columns in Table 5. As shown in Table 5, there are 16 connected vectors marked "Y" in column "Connected?", and 14 of them are feasible vectors, marked as "Y" in column "Feasible?". The reliability of each feasible vector is presented in the last column of Table 5.

Table 5. All obtained 5-tuple vectors using the BAT.

| $i$ | $X_i$ | $C(X_i)$ | Connected? | Feasible? | $R(G(X_i))$ | $i$ | $X_i$ | $C(X_i)$ | Connected? | Feasible? | $R(G(X_i))$ |
|---|---|---|---|---|---|---|---|---|---|---|---|
| 1 | (0, 0, 0, 0, 0) | 0 | | | | 17 | (0, 0, 0, 0, 1) | 10 | | | |
| 2 | (1, 0, 0, 0, 0) | 2 | | | | 18 | (1, 0, 0, 0, 1) | 12 | | | |
| 3 | (0, 1, 0, 0, 0) | 4 | | | | 19 | (0, 1, 0, 0, 1) | 14 | Y | Y | 0.67500 |
| 4 | (1, 1, 0, 0, 0) | 6 | | | | 20 | (1, 1, 0, 0, 1) | 16 | Y | Y | 0.67500 |
| 5 | (0, 0, 1, 0, 0) | 6 | | | | 21 | (0, 0, 1, 0, 1) | 16 | | | |

| | | | | | | | | | |
|---|---|---|---|---|---|---|---|---|---|
| 6 (1, 0, 1, 0, 0) | 8 | | | | 22 (1, 0, 1, 0, 1) | 18 | Y | Y | 0.60562 |
| 7 (0, 1, 1, 0, 0) | 10 | | | | 23 (0, 1, 1, 0, 1) | 20 | Y | Y | 0.67500 |
| 8 (1, 1, 1, 0, 0) | 12 | | | | 24 (1, 1, 1, 0, 1) | 22 | Y | Y | 0.73556 |
| 9 (0, 0, 0, 1, 0) | 7 | | | | 25 (0, 0, 0, 1, 1) | 17 | | | |
| 10 (1, 0, 0, 1, 0) | 9 | Y | Y | 0.76000 | 26 (1, 0, 0, 1, 1) | 19 | Y | Y | 0.67500 |
| 11 (0, 1, 0, 1, 0) | 11 | | | | 27 (0, 1, 0, 1, 1) | 21 | Y | Y | 0.67500 |
| 12 (1, 1, 0, 1, 0) | 13 | Y | Y | 0.76000 | 28 (1, 1, 0, 1, 1) | 23 | Y | Y | **0.92200** |
| 13 (0, 0, 1, 1, 0) | 13 | | | | 29 (0, 0, 1, 1, 1) | 23 | | | |
| 14 (1, 0, 1, 1, 0) | 15 | Y | Y | 0.76000 | 30 (1, 0, 1, 1, 1) | 25 | Y | Y | 0.88113 |
| 15 (0, 1, 1, 1, 0) | 17 | Y | Y | 0.61200 | 31 (0, 1, 1, 1, 1) | 27 | Y | | |
| 16 (1, 1, 1, 1, 0) | 19 | Y | Y | 0.79060 | 32 (1, 1, 1, 1, 1) | 29 | Y | | |

From the above, $X_{28}$ = (1, 1, 0, 1, 1) has the best reliability $R(X_{28})$ = 0.92200 with $C(X_{28})$ = 23 < 26. Hence, the most reliable network is shown in Figure 2 if the budget limit is 26.

Compared with the traditional MP-based problem [15, 17, 28, 29, 30], which only limits the budget of each MP, the proposed budget-based binary-state network reliability can consider the entire network and is more reasonable and practical for current real-life networks. In the rest of the study, we focus on finding the most reliable network that satisfies the budget limit efficiently.

## 5. PROPOSED NOVEL CONCEPTS

To find the most reliable network that satisfies the budget limit in the proposed novel budget-based binary-state network reliability problem effectively, the number of infeasible and feasible vectors that are optimal must both be reduced. Three major innovations, including the dominant rule, maximal/minimal number of arcs in networks, and stepwise BAT for finding feasible vectors, are proposed to achieve the above goal, and are discussed in this section.

### 5.1 Dominance Rule

The proposed algorithm is based on BAT. To improve the efficiency of the BAT, a dominance rule is proposed to reduce vectors and all related vectors based on the following important property:

$$\text{If } G(X) \subseteq G(Y), \text{ we have } X \leq Y, \Pr(X) \leq \Pr(Y), \text{ and } C(X) \leq C(Y). \tag{7}$$

Let $X$ be a vector, $X^{\subseteq}$ = {vector $X^*$ | for all $X^*$ with $X \subseteq X^*$}, and $X^{\supseteq}$ = {vector $X^*$ | for all $X^* \subseteq X$}.

For example, in Table 5, $X_{31}^{\subseteq}$ = {$X_{31}$ = (0, 1, 1, 1, 1), $X_{32}$ = (1, 1, 1, 1, 1)} and $X_{16}^{\supseteq}$ = {$X_{16}$ = (1, 1, 1,

1, 0), $X_1 = (0, 0, 0, 0, 0)$, $X_2 = (1, 0, 0, 0, 0)$, …, $X_{15} = (0, 1, 1, 1, 0)$ }.

Based on Eq. (7), we have

1. All vectors in $X^\subseteq$ can be discarded if $X$ is infeasible, i.e., $C^* \leq C(X)$.

2. All vectors in $X^\supseteq$ can be discarded if $R(X) \leq R(X^*)$ for at least one feasible vector $X^*$.

For example, in Table 5, all vectors in $X_{31}^\subseteq = \{X_{31}, X_{32}\}$ can be discarded because $C(X_{31}) = 27 > 26$, i.e., the cost of each vector in $X_{31}^\subseteq$ is larger than the budget and is therefore infeasible. Similarly, all vectors in $X_{16}^\supseteq = \{X_{16}, X_1, X_2, …, X_{15}\}$ are not optimal and can be discarded because $R(X_{16}) = 0.7906 < R(X_{28}) = 0.922$. Note that $G(X_{28}) = G(1, 1, 0, 1, 1)$ is the most reliable network, with $R(X_{28}) = 0.922$, as illustrated in Figure 1.

## 5.2 $m_{max}$ and $m_{min}$

Let $m_{max}$ and $m_{min}$ be numbers such that $m_{min} \leq (x_1 + x_2 + … + x_m) \leq m_{max}$ for all feasible $X = (x_1, x_2, …, x_m)$, respectively, and $C(a_{[k]}) < C(a_{[k+1]})$ for all $a_{[k]} \in E$ and $k = 1, 2, …, m-1$. We have that $m_{max}$ is the largest $i$ such that

$$\sum_{k=1}^{i} C(a_{[k]}) \leq C^* \leq \sum_{k=j}^{m} C(a_{[k]}), \tag{8}$$

and

$$m_{min} = \text{Min } \{|p| \mid \text{for all MP } p \text{ with } C(p) \leq C^*\}. \tag{9}$$

Hence, each feasible vector $X$ must have $m_{min} \leq \sum_{k=1}^{m} X(a_k) \leq m_{max}$. This important property is implemented in the proposed BAT without wasting time in finding infeasible vector $X$ with $\sum_{k=1}^{m} X(a_k) < m_{min}$ or $\sum_{k=1}^{m} X(a_k) > m_{max}$ to reduce the number of infeasible vectors.

For example, in Figure 1 and Table 3, because $a_k = a_{[k]}$ for all $k = 1, 2, …, m$, $C(a_1) = 2 < C(a_2)$

$= 4 < C(a_3) = 6 < C(a_4) = 7 < C(a_5) = 10$. We have $m_{max} = m_{min} = 4$ for all feasible vectors because

$$\sum_{k=1}^{4} C(a_k) < C^* = 27 < \sum_{k=1}^{5} C(a_k) = 29 \text{ and } \sum_{k=3}^{5} C(a_k) = 23 < C^* = 27 = \sum_{k=2}^{5} C(a_k),$$ respectively. Note that $X_{28} = (1, 1, 0, 1, 1)$ is the best solution, and that there are exactly four arcs in $G(X_{28})$ in Figure 1.

## 5.3 $B_d^=$, $X_d^*$, and $X_d^\#$

Let $B_d^= = \{X = (x_1, x_2, \ldots, x_m)|$ for all vectors $X$ obtained from BAT with $\sum_{k=1}^{m} C(a_k) = d$ and $X$ is not less than $X\hat{\ }$ for all feasible $X\hat{\ } \in B_{d*}^=$ and $d < d^* \leq m_{max}\}$. $X_d^*$ and $X_d^\#$ are the best feasible vectors in $\{X |$ for all feasible $X$ in $B_d^=\}$ and vector in $B_d^=$, respectively. Hence,

$$C(X_d^*) \leq \text{Min}\{C(X_d^\#), C^*\}, R(X) \leq R(X_d^*), \text{ and } R(Y) \leq R(X_d^\#) \tag{10}$$

for each feasible vector $X$ and all vectors $Y$ in $B_d^=$, respectively.

For example, in Figure 1, $B_{m_{max}}^= = B_4^= = \{X_{16} = (1, 1, 1, 1, 0), X_{24} = (1, 1, 1, 0, 1), X_{28} = (1, 1, 0, 1, 1), X_{30} = (1, 0, 1, 1, 1), X_{31} = (0, 1, 1, 1, 1)\}$ and $B_3^= = \emptyset$.

In Section 5.2, the best feasible solution for the proposed problem is $B_d^=$ for all $d = m_{min}, m_{min}+1, \ldots, m_{max}$, i.e., the best feasible vector in $\{X_d^* | d = m_{min}, m_{min}+1, \ldots, m_{max}\}$ is the best solution for the proposed problem.

Also, in Section 5.1, $X_{m_{max}}^*$ is the best solution to the proposed problem if $X_{m_{max}}^* = X_{m_{max}}^\#$. For example, in Figure 1, $X_{28} = (1, 1, 0, 1, 1)$ with $R(X_{28}) = 0.922$ is the best feasible vector among all vectors in $B_{m_{max}}^=$, i.e., $X_{28} = X_4^* = X_4^\#$. Hence, $X_{28}$ is also the best feasible solution with the best reliability, $R(X_{28}) = 0.922$, and $C(X_{28}) = 23 < C^* = 27$.

However, if $X_{m_{max}}^* \neq X_{m_{max}}^\#$, $X_{m_{max}}^*$ is still the best feasible vector among all feasible vectors in $B_{m_{max}}^=$, but may not be the best feasible solution for the problem. Hence, we must find $\delta$ until either

$X_\delta^* = X_\delta^\#$ or $B_\delta^= = \emptyset$ for $d = m_{max}-1, m_{max}-2, \ldots, m_{min}$. The best solution in $\{ X_d^* \mid d = m_{min}, m_{min}+1, \ldots, \delta\}$ is also the most feasible solution.

### 5.4 Stepwise Vector

From the last paragraph of Section 5.2, we must find all vectors in $B_d^=$ for $d = m_{max}, m_{max}-1, m_{max}-2, \ldots, m_{min}$. The traditional BAT finds all binary-state vectors, including those that are connected and disconnected. Hence, to simplify the traditional BAT without requiring all infeasible vectors, a new vector called a stepwise vector is proposed such that all found vectors are only in $B_d^=$ for $d = m_{max}, m_{max}-1, m_{max}-2, \ldots, m_{min}$.

A stepwise vector $X$ is a special vector such that $x_1 < x_2 < \ldots < x_d$ and $X = (x_1, x_2, \ldots, x_d) \leq U = (m-d+1, \ldots, m-1, m)$. Each stepwise vector $X = (x_1, x_2, \ldots, x_d)$ can be represented in binary-state form

$$B(X) = (b_1, b_2, \ldots, b_m) \text{ and } b_i = \begin{cases} 1 & x_i = j \\ 0 & \text{otherwise} \end{cases} \tag{11}$$

for $i = 1, 2\ldots, d$ and $j = 1, 2, \ldots, m$.

For example, in Figure 1, each stepwise vector $X$ and its $B(X)$ are provided in Table 6. Note that there is no need to find $B_1^=$, $B_2^=$, and $B_5^=$ if $d = 2, 3, 4$, and the number of vectors is reduced from 32 to 27 after using stepwise vectors.

**Table 6.** Stepwise vector $X$ and $B(X)$ in $B_d^=$ for $d=1, 2, \ldots, 5$, based on Fig. 1.

| $i$ | $X_i \in B_1^=$ | $B(X)$ | $X_i \in B_2^=$ | $B(X)$ | $X_i \in B_3^=$ | $B(X)$ | $X_i \in B_4^=$ | $B(X)$ | $X_i \in B_5^=$ | $B(X)$ |
|---|---|---|---|---|---|---|---|---|---|---|
| 1 | (1) | (1, 0, 0, 0, 0) | (1, 2) | (1, 1, 0, 0, 0) | (1, 2, 3) | (1, 1, 1, 0, 0) | (1, 2, 3, 4) | (1, 1, 1, 1, 0) | (1, 2, 3, 4, 5) | (1, 1, 1, 1, 1) |
| 2 | (2) | (0, 1, 0, 0, 0) | (1, 3) | (1, 0, 1, 0, 0) | (1, 2, 4) | (1, 1, 0, 1, 0) | (1, 2, 3, 5) | (1, 1, 1, 0, 1) | | |
| 3 | (3) | (0, 0, 1, 0, 0) | (1, 4) | (1, 0, 0, 1, 0) | (1, 2, 5) | (1, 1, 0, 0, 1) | (1, 2, 4, 5) | (1, 1, 0, 1, 1) | | |
| 4 | (4) | (0, 0, 0, 1, 0) | (1, 5) | (1, 0, 0, 0, 1) | (1, 3, 4) | (1, 0, 1, 1, 0) | (1, 3, 4, 5) | (1, 0, 1, 1, 1) | | |
| 5 | (5) | (0, 0, 0, 0, 1) | (2, 3) | (0, 1, 1, 0, 0) | (1, 3, 5) | (1, 0, 1, 0, 1) | (2, 3, 4, 5) | (0, 1, 1, 1, 1) | | |
| 6 | | | (2, 4) | (0, 1, 0, 1, 0) | (1, 4, 5) | (1, 0, 0, 1, 1) | | | | |
| 7 | | | (2, 5) | (0, 1, 0, 0, 1) | (2, 3, 4) | (0, 1, 1, 1, 0) | | | | |
| 8 | | | (3, 4) | (0, 0, 1, 1, 0) | (2, 3, 5) | (0, 1, 1, 0, 1) | | | | |
| 9 | | | (3, 5) | (0, 0, 1, 0, 1) | (2, 4, 5) | (0, 1, 0, 1, 1) | | | | |
| 10 | | | (4, 5) | (0, 0, 0, 1, 1) | (3, 4, 5) | (0, 0, 1, 1, 1) | | | | |

### 5.5 Stepwise BAT and Example

From Section 5.2, it is better to find stepwise vectors rather than all binary-state vectors because the number of all stepwise vectors and coordinates are both less than those of the binary-state vectors. In this section, a novel BAT modified from the BAT [37] is proposed to find all stepwise vectors together with its pseudocode, time complexity, and a demonstrated example.

### 5.5.1 Pseudocode

The pseudocode for the proposed stepwise BAT is represented in the following.

**Algorithm: Stepwise BAT**

**Input:**   $m$, $\mu$, and $\mu < m$.

**Output:**  All stepwise vectors $X$ in $B_d^=$ for $d = m_{max}, m_{max}-1, m_{max}-2, \ldots, m_{min}$.

**STEP 0.** Let $i = \mu$, $U = (m-\mu+1, \ldots, m-1, m)$, $S = (1, 2, \ldots, \mu)$, $X^* = \mathbf{0}$, $R(X^*) = 0$, and go to STEP 2.

**STEP 1.** If $s_i < u_i$, let $s_i = s_i + 1$, $s_j = s_{j-1} + 1$ for $j = i+1, i+2, \ldots, \mu$, and $i = \mu$. Otherwise, go to STEP 3.

**STEP 2.** If $X = B(S)$ is connected, $C(X) > C^*$, and $R(X^*) < R(X)$, let $X^* = X$. Go to STEP 1.

**STEP 3.** If $i = 1$, halt.

**STEP 4.** Let $i = i - 1$ and go to STEP 1.

Because

$$s_1 < s_2 < \ldots < s_\mu \text{ and } S = (s_1, s_2, \ldots, s_\mu) \leq U = (m-\mu+1, \ldots, m-1, m) \text{ for all } S, \qquad (12)$$

the pseudocode above finds all stepwise vectors. STEP 2 verifies whether these stepwise vectors are connected, feasible, and have better reliability than the current $R(X^*)$. Hence, all feasible connected stepwise vectors are filtered out, and the one with the best reliability is obtained in STEP 2.

Moreover, the number of all $\mu$-tuple stepwise vectors is $C_\mu^m = \dfrac{m!}{(m-\mu)!\mu!}$, all stepwise vectors are $O(2^m)$, the time complexity to verify the connectivity using PLSA [37, 42], feasibility using Eq. (12), and calculated reliability of $B(X)$ based on BAT are $O(n)$, $O(n)$, and $m2^m$ [37, 38], respectively. Hence, the proposed stepwise BAT can solve the proposed problem correctly with a time complexity $O(m2^{2m})$.

### 5.5.2 Step-by-Step Example

Let $m = 5$, $\mu = 3$, $\Pr(a_i)$ is listed in Table 3, and $C(a_i)$ is listed in Table 7 under setting A, where $i = 1, 2, \ldots, 5$.

Table 7. Cost of each arc under setting A, B, and C in Figure 1, respectively.

|   | $C(a_1)$ | $C(a_2)$ | $C(a_3)$ | $C(a_4)$ | $C(a_5)$ |
|---|---|---|---|---|---|
| A | 10 | 15 | 20 | 25 | 30 |
| B | 30 | 25 | 20 | 15 | 10 |
| C | 15 | 20 | 10 | 25 | 30 |

Based on the above pseudocode of the proposed stepwise BAT, all connected feasible stepwise vector $X$ can be found in Figure 1 after letting as listed below.

**STEP 0.** Let $i = 3$, $U = (5-3+1=3, 4, 5)$, $S = (1, 2, 3)$, and go to STEP 2.

**STEP 2.** Because $B(S) = (1, 1, 1, 0, 0)$ is disconnected, go to STEP 1.

**STEP 1.** Because $s_3 = 3 < u_3 = 5$, let $s_3 = s_3 + 1 = 4$, i.e., $S = (1, 2, 4)$, and $i = 3$.

**STEP 2.** Because $B(S) = (1, 1, 0, 1, 0)$ is disconnected, go to STEP 1.

**STEP 1.** Because $s_3 = 4 < u_3 = 5$, let $s_3 = s_3 + 1 = 5$, i.e., $S = (1, 2, 5)$, and $i = 3$.

**STEP 2.** Because $B(S) = (1, 1, 0, 0, 1)$ is disconnected, go to STEP 1.

**STEP 1.** Because $s_3 = u_3 = 5$, go to STEP 3.

**STEP 3.** Because $i = 3$, go to STEP 4.

**STEP 4.** Let $i = i - 1 = 2$ and go to STEP 1.

**STEP 1.** Because $s_2 = 2 < u_2 = 4$, let $s_2 = s_2 + 1 = 3$, $s_3 = s_2 + 1 = 4$, i.e., $X = (1, 3, 4)$, and $i = 3$.

**STEP 2.** Because $B(X) = (1, 0, 1, 1, 0)$ is disconnected, go to STEP 1.

**STEP 1.** Because $s_3 = 4 < u_3 = 5$, let $s_3 = s_3 + 1 = 5$, i.e., $X = (1, 3, 5)$, and $i = 3$.

**STEP 2.** Because $B(X) = (1, 0, 1, 0, 1)$ is connected, $C(a_1) + C(a_3) + C(a_5) = 60 < C^* = 75$, $R(X) = 0.60562 > R_{max} = 0$, let $R_{max} = R(X) = 0.60562$. Go to STEP 1.

**STEP 1.** Because $s_3 = u_3 = 5$, let $s_3 = s_3 + 1 = 5$, i.e., $S = (1, 3, 5)$, and $i = 3$.

**STEP 3.** Because $i = 3$, go to STEP 4.

**STEP 4.** Let $i = i - 1 = 2$ and go to STEP 1.

**STEP 1.** Because $s_2 = 3 < u_2 = 4$, let $s_2 = s_2 + 1 = 4$, $s_3 = s_2 + 1 = 5$, i.e., $X = (1, 4, 5)$, and $i = 3$.

$$\vdots$$

$$\vdots$$

Let the budget limit $C^* = 77$. The complete procedure is provided in Table 8. In Table 8, all strikethrough values are over $C^* = 77$ and are infeasible; for example, $C(X_1) = 100$ is infeasible in $B_d^=$ for $d = 5$; all $X$ in these empty cells are disconnected after verification by PLSA [37, 42]; each best solution is marked * in the related values of A, B, and C.

**Table 8.** Stepwise vector $X$ and $B(X)$ in $B_d^=$ for $d$=2, 3, 4, 5, based on Figure 1 and Table 7.

| d | i | X | B(X) | R(X) | C(X) A | B | C |
|---|---|---|---|---|---|---|---|
| 2 | 1 | (1, 2) | (1, 1, 0, 0, 0) | | | | |
|   | 2 | (1, 3) | (1, 0, 1, 0, 0) | | | | |
|   | 3 | (1, 4) | (1, 0, 0, 1, 0) | 0.7600000 | 35 | 45 | 40 |
|   | 4 | (1, 5) | (1, 0, 0, 0, 1) | | | | |
|   | 5 | (2, 3) | (0, 1, 1, 0, 0) | | | | |
|   | 6 | (2, 4) | (0, 1, 0, 1, 0) | | | | |
|   | 7 | (2, 5) | (0, 1, 0, 0, 1) | 0.6750000 | 45 | 35 | 50 |
|   | 8 | (3, 4) | (0, 0, 1, 1, 0) | | | | |
|   | 9 | (3, 5) | (0, 0, 1, 0, 1) | | | | |
|   | 10 | (4, 5) | (0, 0, 0, 1, 1) | | | | |
| 3 | 1 | (1, 2, 3) | (1, 1, 1, 0, 0) | | | | |
|   | 2 | (1, 2, 4) | (1, 1, 0, 1, 0) | | | | |
|   | 3 | (1, 2, 5) | (1, 1, 0, 0, 1) | | | | |
|   | 4 | (1, 3, 4) | (1, 0, 1, 1, 0) | | | | |
|   | 5 | (1, 3, 5) | (1, 0, 1, 0, 1) | 0.6056250 | 60 | 60 | 55 |
|   | 6 | (1, 4, 5) | (1, 0, 0, 1, 1) | | | | |
|   | 7 | (2, 3, 4) | (0, 1, 1, 1, 0) | 0.6120000 | 60 | 60 | 55 |
|   | 8 | (2, 3, 5) | (0, 1, 1, 0, 1) | | | | |
|   | 9 | (2, 4, 5) | (0, 1, 0, 1, 1) | | | | |
|   | 10 | (3, 4, 5) | (0, 0, 1, 1, 1) | | | | |
| 4 | 1 | (1, 2, 3, 4) | (1, 1, 1, 1, 0) | 0.7906000 | 70* | ~~90~~ | 70* |
|   | 2 | (1, 2, 3, 5) | (1, 1, 1, 0, 1) | 0.7355625 | 75 | ~~85~~ | 75 |
|   | 3 | (1, 2, 4, 5) | (1, 1, 0, 1, 1) | | ~~80~~ | ~~80~~ | ~~90~~ |
|   | 4 | (1, 3, 4, 5) | (1, 0, 1, 1, 1) | 0.8811250 | ~~85~~ | 75* | ~~80~~ |
|   | 5 | (2, 3, 4, 5) | (0, 1, 1, 1, 1) | 0.8280000 | ~~90~~ | 70 | ~~85~~ |
| 5 | 1 | (1, 2, 3, 4, 5) | (1, 1, 1, 1, 1) | | ~~100~~ | ~~100~~ | ~~100~~ |

## 6. PROPOSED ALGORITHM

The proposed BAT-based algorithm for solving the novel budget-based binary-state network reliability problem is introduced formally in this section by combining all novelties proposed in Section 5.

### 6.1 Pseudocode

From Section 5.1, $X$ is the optimal solution if it is also the best vector in the $B^=_{m_{max}}$. Hence, it is better to find all $B^=_d$ for $i = m_{max}, m_{max} -1, \ldots, m_{min}$, i.e., from larger $d$ to smaller $d$. Based on the above property and other properties found in Section 5, the pseudocode of the proposed bottom-up stepwise BAT is listed as follows.

**Algorithm 5: New Bottom-Up Stepwise BAT**

**Input:** $G(V, E, D_B)$, $C^*$, source node 1 and sink node $n$.

**Output:** The most reliable $G(X) \subseteq G(V, E)$ satisfied the budget-limit.

**STEP 0.** Find $m_{max}$ based on Eq. (9), let $d = m_{max}$, $X^* = \mathbf{0}$, and $R(X^*) = 0$.

**STEP 1.** Find $B^=_d$ and the best solution $X^*_d$ in $B^=_d$ using the proposed stepwise BAT.

**STEP 2.** If $B^=_d = \emptyset$, go to STEP 6.

**STEP 3.** If $R(X^*) < R(X^*_d)$, let $X^* = X^*_d$.

**STEP 4.** If all vectors in $B^=_d$ are feasible, halt and $X^*$ is the best vector.

**STEP 5.** If $R(X) \leq R(X^*_d)$ for all $X$ in $B^=_d$, halt and $X^*$ is the best vector.

**STEP 6.** Let $d = d - 1$ and go to STEP 1.

In the pseudocode, STEP 0 initializes the values of $m_{max}$, $d$, $X^*$ and $R(X^*)$. STEP 1 implements the stepwise BAT proposed in Section 5.5 to find $B^=_d$ and $X^*_d$. STEPs 2-5 determine whether $B^=_d$ is empty, has a better vector, or all vectors are feasible, respectively. If $B^=_d$ is empty in STEP 2, let $d =$ $d-1$ and proceed to STEP 6 for the next $B_d^=$. The best solution $X_d^*$ in $B_d^=$ replaces the current best solution $X^*$ if $X_d^*$ is better, as stated in STEP 3. Halt, if all solutions are feasible in $B_d^=$ in STEP 4 or $R(X) \leq R(X_d^*)$ for all $X$ in $B_d^=$ in STEP 5. The PLSA [42] is implemented to verify the connectivity of each vector in $B_d^=$.

Finding $B_d^=$ is the major part of the time complexity of the proposed bottom-up stepwise BAT algorithm. As shown in Section 5.5, $O(m C_d^m)$ is required to verify the feasibility of each vector in $B_d^=$. Because

$$O(m \sum_{d=m_{\min}}^{m_{\max}} C_d^m) \leq O(m \sum_{d=0}^{m} C_d^m) = O(m2^m), \tag{13}$$

which is the time complexity in finding all MPs in the traditional MP-based binary-state network reliability with a budget-limit problem [15, 17, 28, 29, 30], the proposed problem is more complicated than the traditional MP-based binary-state network reliability with a budget-limit problem. However, the proposed bottom-up stepwise BAT is more efficient than the MP algorithm in solving the proposed problem.

## 6.2 Example

Solving the MP-based budget-limit binary state network reliability problem is a two-fold problem, NP-hard and #P-hard [12, 13], because it requires searching for all feasible MPs that satisfy the budget limit and calculating the exact binary-state network reliability based on feasible MPs in binary-state networks [15, 17, 28, 29, 30].

The traditional MP-based budget-limit binary-state network reliability problem only considers the budget limit for each MP [15, 17, 28, 29, 30], which is less complicated than the proposed new problem in considering the budget of all arcs in building a reliable network. Hence, to make readers understand the proposed novel problem and novel algorithm easily and quickly, Figure 1, adapted from the popular benchmark binary-state networks with a suitable size [15, 17, 28, 29, 30], is implemented to demonstrate the proposed algorithm step by step.

For readers to easily understand the proposed problem and algorithm, four different cost limits: $C_1^* = 110$, $C_2^* = 95$, $C_3^* = 85$, $C_4^* = 65$, and $C_5^* = 40$ are considered here. Let the cost of each arc be the same as for setting A, as listed in Table 7. The procedure in finding the best vector $X^*$ for four different cost limits $C_1^* = 110$, $C_2^* = 95$, $C_3^* = 85$, $C_4^* = 65$, and $C_5^* = 40$ are listed in Tables 9-13, respectively.

**Table 9.** Procedure in finding $X^*$ for $C_1^* = 110$.

| $d$ | $i$ | $X_i$ | $B(X_i)$ | $R(X_i)$ | $C(X_i)$ | Remark |
|---|---|---|---|---|---|---|
| $d_{max}$=5 | 1 | (1, 2, 3, 4, 5) | (1, 1, 1, 1, 1) | **0.9417625** | 100 | By STEP 4 |

**Table 10.** Procedure in finding $X^*$ for $C_2^* = 95$.

| $d$ | $i$ | $X_i$ | $B(X_i)$ | $R(X_i)$ | $C(X_i)$ | Remark |
|---|---|---|---|---|---|---|
| $d_{max}$=4 | 1 | (1, 2, 3, 4) | (1, 1, 1, 1, 0) | 0.7906000 | 70 | |
| | 2 | (1, 2, 3, 5) | (1, 1, 1, 0, 1) | 0.7355625 | 75 | |
| | 3 | (1, 2, 4, 5) | (1, 1, 0, 1, 1) | **0.9220000** | 80 | By STEP 4 |
| | 4 | (1, 3, 4, 5) | (1, 0, 1, 1, 1) | 0.8811250 | 85 | |
| | 5 | (2, 3, 4, 5) | (0, 1, 1, 1, 1) | 0.8280000 | 90 | |

**Table 11.** Procedure in finding $X^*$ for $C_3^* = 85$.

| $d$ | $i$ | $X_i$ | $B(X_i)$ | $R(X_i)$ | $C(X_i)$ | Remark |
|---|---|---|---|---|---|---|
| $d_{max}$=4 | 1 | (1, 2, 3, 4) | (1, 1, 1, 1, 0) | 0.7906000 | 70 | |
| | 2 | (1, 2, 3, 5) | (1, 1, 1, 0, 1) | 0.7355625 | 75 | |
| | 3 | (1, 2, 4, 5) | (1, 1, 0, 1, 1) | **0.9220000** | 80 | By STEP 5 |
| | 4 | (1, 3, 4, 5) | (1, 0, 1, 1, 1) | 0.8811250 | 85 | |
| | 5 | (2, 3, 4, 5) | (0, 1, 1, 1, 1) | 0.8280000 | 90 | $C(X_i) > C_3^* = 85$ |

**Table 12.** Procedure in finding $X^*$ for $C_4^* = 65$.

| $d$ | $i$ | $X_i$ | $B(X_i)$ | $R(X_i)$ | $C(X_i)$ | Remark |
|---|---|---|---|---|---|---|
| $d_{max}$=3 | 1 | (1, 2, 3) | (1, 1, 1, 0, 0) | | | disconnect |
| | 2 | (1, 2, 4) | (1, 1, 0, 1, 0) | **0.7600000** | 50 | By STEP 5 |
| | 3 | (1, 2, 5) | (1, 1, 0, 0, 1) | 0.6750000 | 55 | |
| | 4 | (1, 3, 4) | (1, 0, 1, 1, 0) | 0.7600000 | 55 | |
| | 5 | (1, 3, 5) | (1, 0, 1, 0, 1) | 0.6056250 | 60 | |
| | 6 | (1, 4, 5) | (1, 0, 0, 1, 1) | 0.7600000 | 65 | |
| | 7 | (2, 3, 4) | (0, 1, 1, 1, 0) | 0.6120000 | 60 | |
| | 8 | (2, 3, 5) | (0, 1, 1, 0, 1) | 0.6750000 | 65 | |
| | 9 | (2, 4, 5) | (0, 1, 0, 1, 1) | 0.6750000 | 70 | $C(X_i) > C_4^* = 65$ |
| | 10 | (3, 4, 5) | (0, 0, 1, 1, 1) | | | disconnect |

Table 13. Procedure in finding $X^*$ for $C_5^* = 40$, where $d_{max}=d_{min}=2$

| $d$ | $i$ | $X_i$ | $B(X_i)$ | $R(X_i)$ | $C(X_i)$ | |
|---|---|---|---|---|---|---|
| $d_{max}=2$ | 1 | (1, 2) | (1, 1, 0, 0, 0) | | | disconnect |
| | 2 | (1, 3) | (1, 0, 1, 0, 0) | | | disconnect |
| | 3 | (1, 4) | (1, 0, 0, 1, 0) | **0.7600000** | 35 | By STEP 5 |
| | 4 | (1, 5) | (1, 0, 0, 0, 1) | | | disconnect |
| | 5 | (2, 3) | (0, 1, 1, 0, 0) | | | disconnect |
| | 6 | (2, 4) | (0, 1, 0, 1, 0) | | | disconnect |
| | 7 | (2, 5) | (0, 1, 0, 0, 1) | 0.6750000 | 45 | $C(X_i)>C_4^*=65$ |
| | 8 | (3, 4) | (0, 0, 1, 1, 0) | | | disconnect |
| | 9 | (3, 5) | (0, 0, 1, 0, 1) | | | disconnect |
| | 10 | (4, 5) | (0, 0, 0, 1, 1) | | | disconnect |

From Tables 9-13, the higher the cost limits, the smaller the number of feasible vectors that need to be found, and the lower the runtime required to obtain the best solution, and vice versa. The total number of feasible vectors depends on the cost limit. However, the number of connected vectors is always fixed.

**6.3 Computation Experiments**

To validate the performance of the proposed algorithm, an experiment was conducted on a real-life 24-pipelines-and-15-town water distribution network located in New Taipei city of Taiwan, as shown in Figure 3 [43] was implemented to investigate the proposed algorithm. We want to build the most reliable water distribution network with a limited budget. The reliability and cost of each arc are presented in Table 14.

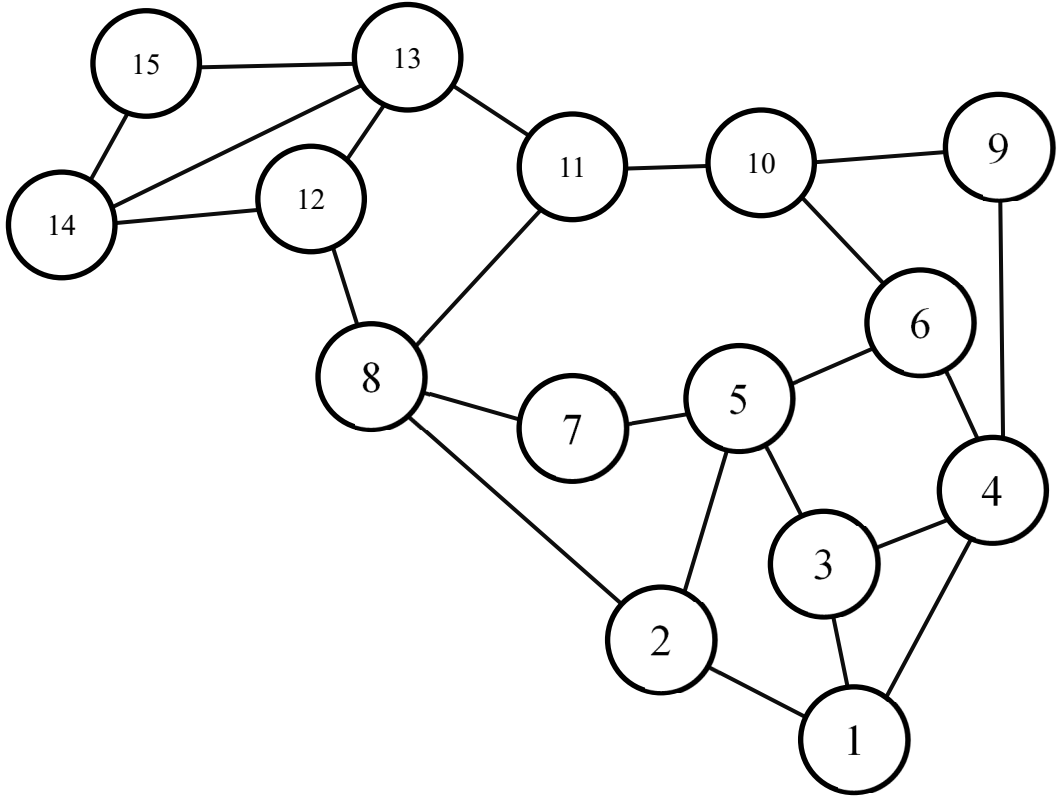

**Figure 3.** Water distribution network.

All MP-based algorithms are only applicable to MP-based budget-limited problems and not to the proposed budget-limited problems. Hence, the performance of the proposed algorithm was compared with that of an algorithm based on BAT. For a fair comparison, the tests for both algorithms were performed under the same computer environments.

**Table 14.** State distribution of each arc in Figure 3.

| $i$ | $a_i=e$ | Pr($e$) | C($e$) | $i$ | $a_i=e$ | Pr($e$) | C($e$) | $i$ | $a_i=e$ | Pr($e$) | C($e$) |
|---|---|---|---|---|---|---|---|---|---|---|---|
| 1 | $e_{1,2}$ | 0.8012 | 69 | 9 | $e_{4,9}$ | 0.8450 | 95 | 17 | $e_{10,11}$ | 0.8680 | 75 |
| 2 | $e_{1,3}$ | 0.9604 | 100 | 10 | $e_{5,6}$ | 0.8506 | 78 | 18 | $e_{11,13}$ | 0.9932 | 79 |
| 3 | $e_{1,4}$ | 0.9811 | 99 | 11 | $e_{5,7}$ | 0.8755 | 79 | 19 | $e_{12,13}$ | 0.9983 | 79 |
| 4 | $e_{2,5}$ | 0.8430 | 78 | 12 | $e_{6,10}$ | 0.9582 | 74 | 20 | $e_{12,14}$ | 0.9557 | 70 |
| 5 | $e_{2,8}$ | 0.8103 | 89 | 13 | $e_{7,8}$ | 0.8826 | 74 | 21 | $e_{13,14}$ | 0.8928 | 70 |
| 6 | $e_{3,4}$ | 0.8061 | 88 | 14 | $e_{8,11}$ | 0.8129 | 91 | 22 | $e_{13,15}$ | 0.8265 | 84 |
| 7 | $e_{3,5}$ | 0.8814 | 98 | 15 | $e_{8,12}$ | 0.8944 | 96 | 23 | $e_{14,15}$ | 0.8576 | 99 |
| 8 | $e_{4,6}$ | 0.8252 | 72 | 16 | $e_{9,10}$ | 0.9274 | 99 | | | | |

Table 15 lists the experimental results from both algorithms. The notations $N$, $T_\bullet$, and $X^*$ are the numbers of all feasible vectors, the runtime for algorithm $\bullet$, and the optimal feasible vector, respectively. Each best runtime between the proposed algorithm and the traditional BAT algorithm is denoted in bold.

From Table 15, the larger $N^*$, the more efficient the proposed algorithm than BAT. The reason is that the proposed bottom-up stepwise BAT finds and verifies only feasible stepwise vectors in $B_d^=$ for $d = m_{max}, m_{max}–1, m_{max}–2, …, 1$, and $|X^*|$ is always closed to $m_{max}$ as shown in Table 11, where $X^*$ is the optimal solution. However, the BAT always finds all binary-state vectors $B_d^=$ from 1 to $m_{max}$.

For example, only two feasible vectors for $C^* = 420$. The optimal feasible vector $X^*$ is the first one and also the last one found in the proposed algorithm. However, $X^*$ is the second vector found in the BAT. Hence, the runtime of the proposed algorithm is one half of that in the BAT.

**Table 15.** State distribution of Figure 3.

| $C^*$ | $N$ | $T_{new}$ | $T_{BAT}$ | $X^*$ | $R(X^*)$ | $C(X^*)$ | $|X^*|$ | $m_{max}$ |
|---|---|---|---|---|---|---|---|---|
| 420 | 2 | **0.43263** | 0.90800 | (1,0,0,0,1,0,0,0,0,0,0,0,0,1,0,0,0,1,0,0,1,0) | 0.47910 | 100 | 5 | 5 |
| 840 | 40178 | **2.04546** | 4.61300 | (1,1,0,0,1,0,1,0,0,0,1,0,1,0,1,0,0,0,0,1,0,0,1) | 0.64411 | 709 | 9 | 10 |
| 1260 | 1100223 | **28.58889** | 1876.62415 | (1,0,1,1,1,0,0,1,1,1,0,1,0,1,0,1,1,1,0,0,1,1,1) | 0.77923 | 999 | 15 | 15 |
| 1680 | 1528236 | **40.67707** | 3996.00220 | (1,1,1,1,1,0,1,1,0,1,1,1,1,1,1,0,1,1,1,1,1,1,1) | 0.82342 | 1141 | 20 | 20 |
| 1935 | 1529485 | **1.28375** | 4176.72314 | (1,1,1,1,1,1,1,1,1,1,1,1,1,1,1,1,1,1,1,1,1,1,1) | 0.83050 | 1611 | 23 | 23 |

From Table 15, we can see that the proposed algorithm outperforms the traditional BAT in both runtime and number of obtained vectors. In addition, the obtained results, along with the time complexity are discussed in section 6.1.

## 7. CONCLUSIONS

From the example used in Section 5.3, we can determine the important, reasonable, and practical aspects of the proposed problem when considering the cost of the entire network. The proposed algorithm is the first to solve this problem. It is based on a novel stepwise BAT to find all stepwise vectors and to reduce the number of infeasible vectors efficiently. The performance of the proposed algorithm was verified from the experimental results. To establish the advantages of both the proposed problem and algorithm, more real-life applications will be discussed and solved in the future. In future work, the proposed problem and algorithm will also be extended to multistate flow networks.

**insACKNOWLEDGEMENTS**

This research was supported in part by the Ministry of Science and Technology, R.O.C. under grant MOST 107-2221-E-007-072-MY3 and MOST 110-2221-E-007-107-MY3. This article was once submitted to arXiv as a temporary submission that was just for reference and did not provide the copyright.